# Adjusted Haar Wavelet for Application in the Power Systems Disturbance Analysis

Abhisek Ukil

*(Corresponding Author)*

*ABB Corporate Research, Baden, Switzerland*

*Address: Segelhofstrasse 1K, CH-5405, Baden 5 Daettwil, Switzerland*

*Tel: +41 58 586 7034*

*Fax: +41 58 586 7358*

*E-mail:* abhiukil@yahoo.com

Rastko Živanović

*School of Electrical and Electronic Engineering, University of Adelaide, Adelaide, Australia*

*E-mail:* zivanovr@yahoo.com

**Abstract**

Abrupt change detection based on the wavelet transform and threshold method is very effective in detecting the abrupt changes and hence segmenting the signals recorded during disturbances in the electrical power network. The wavelet method estimates the time-instants of the changes in the signal model parameters during the pre-fault condition, after initiation of fault, after circuit-breaker opening and auto-reclosure. Certain kinds of disturbance signals do not show distinct abrupt changes in the signal parameters. In those cases, the standard mother wavelets fail to achieve correct event-specific segmentations. A new adjustment technique to the standard Haar wavelet is proposed in this paper, by introducing 2*n* adjusting zeros in the Haar wavelet scaling filter, *n* being a positive integer. This technique is quite effective in segmenting those fault signals into pre- and post-fault segments, and it is an improvement over the standard mother wavelets for this application. This paper presents many practical examples where recorded signals from the power network in South Africa have been used.

# 1      Introduction

Automatic disturbance recognition and analysis from the recordings of the digital fault recorders (DFRs) play a significant role in fast fault-clearance, helping to achieve a secure and reliable electrical power supply. Segmentation of the fault recordings by detecting the abrupt changes in the characteristics of the fault recordings, obtained from the DFRs of the electrical power network, is the first step towards automatic disturbance recognition and analysis.

To accomplish the abrupt change detection, we propose the use of the wavelet transform to transform the original fault signal into finer wavelet scales, followed by a progressive search for the largest wavelet coefficients on that scale as discussed in [1]. However, for certain kinds of disturbance signals not showing distinct abrupt changes in the signal parameters, standard mother wavelets e.g., Haar (Daubechies 1), Daubechies 4 [2], etc fail to achieve the correct event-specific signal segmentations. So, we propose a new technique for adjustment to the standard Haar wavelet. The new adjusted Haar wavelet can be successfully applied for those disturbance signals, not showing distinct abrupt changes in the signal parameters, to segment them based on the fault-inception time into pre- and post-fault segments.

The remainder of this paper is organized as follows. In Section 2, power systems disturbance analysis as application domain is discussed. Different abrupt change detection-based segmentation techniques relevant to this application are recalled in Section 3. In Section 4, we present the concept of the adjusted Haar wavelet in details. Application results of the adjusted Haar wavelet technique are discussed in Section 5, and conclusions are given in Section 6.

# 2      Power Systems Disturbance Analysis

Automatic analysis of the disturbances in the power transmission network of South Africa depends on the DFR recordings. Presently, 98% of the transmission lines are equipped with the DFRs on the feeder bays, with an additional few installed on the Static Var Compensators (SVCs) and 95% of these are remotely accessible via a X.25 communication system [3]. The DFRs trigger due to reasons like, power network fault conditions, protection operations, breaker operation and the like. Each DFR recording typically consists of 32 points binary information and analog information in the form of voltages and currents per phase as well as the neutral current.

To accomplish an automatic disturbance recognition and analysis, we would first apply the abrupt change detection algorithms to segment the fault recordings into different event-specific segments, e.g., pre-fault segment, after initiation of fault, after circuit-breaker opening, after auto-reclosure of the circuit-breakers. Then we would construct the appropriate feature vectors for the different segments; finally the pattern-matching algorithm would be applied using those feature vectors to accomplish the fault recognition and disturbance analysis tasks. The purpose of this study is to augment the existing semi-automated fault analysis and recognition system with more robust and accurate algorithms and techniques to make it fully automated. The current manual analysis is cumbersome and time-consuming, typically requiring one to 10 hours or more, depending on the complexity and severity of the disturbance event. The complete automatic disturbance recognition and analysis tasks would be performed without any significant human intervention within five minutes of the acquiring of the disturbance signals.

## 3     Abrupt Change Detection-based Segmentation

Detection of abrupt changes in signal characteristics is a much studied subject with many different approaches. It has a significant role to play in failure detection and isolation (FDI) systems and segmentation of signals in recognition oriented signal processing [4].

The current semi-automatic fault analysis system uses peak value detection and superimposed current quantities algorithms [3] for manual segmentation, which are not very accurate. The authors categorized the different automatic abrupt change detection-based segmentation techniques in a comparative manner in [5], which are given as follows.

- *Simple methods*
    - Superimposed Current Quantities
    - Linear Prediction Error Filter

- - Adaptive Whitening Filter
- *Linear Model-based approach*
  - Additive Spectral Changes
  - Auto Regressive (AR) Modeling and Joint Segmentation
  - State-Space Modeling and Recursive Parameter Identification
- *Model-free approach*
  - Support Vector Machines
- *Non-parametric approach*
  - Discrete Fourier Transform
  - Wavelet Transform.

The authors have already established the wavelet transform-based abrupt change detection technique in [1], which is one of the promising and fast methods. Figure 1 shows the result using the wavelet method for the fault signal, sampled at a sampling frequency of 2.5 kHz [3], obtained from the DFRs during a phase-to-ground fault. The wavelet transform uses the Haar (Daubechies 1) [2] mother wavelet.

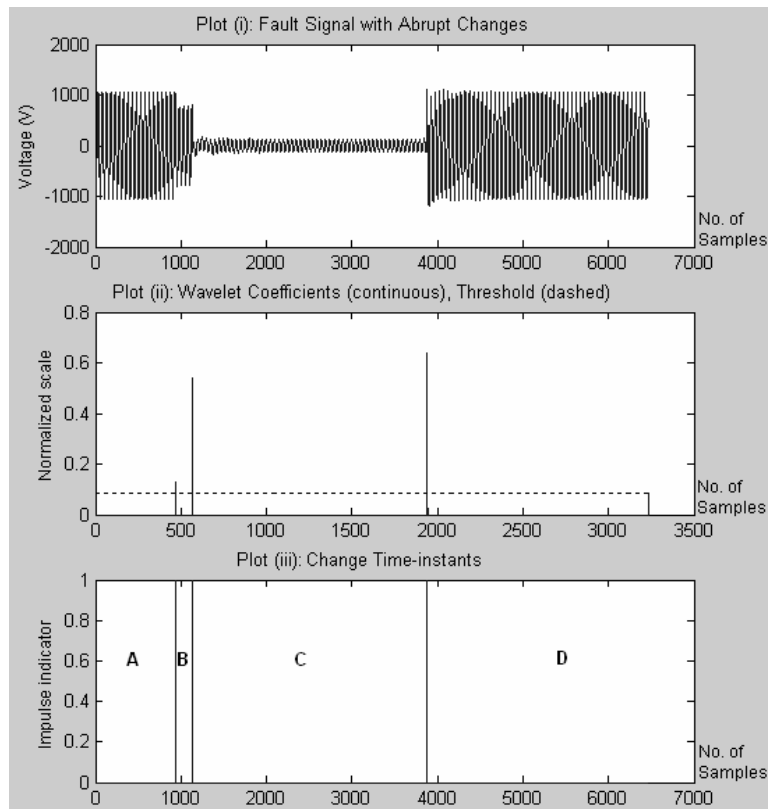

Figure 1.  Abrupt change detection-based segmentation of the BLUE-phase voltage recording during a phase-to-ground fault

In Figure 1, the original DFR recording, from the power transmission network of South Africa, for the voltage during a phase-to-ground fault in the BLUE-Phase, sampled at a frequency of 2.5 kHz [3], is shown in the plot (i); wavelet coefficients for this fault signal and the universal threshold (dashed) are shown in the plot (ii) and the change time-instants as unit impulses computed using the threshold checking (plot ii) followed by the smoothing filtering [1] are shown in the plot (iii). The time-instants of the changes in the signal characteristics in the plot (iii) in Figure 1 indicate the different signal segments owing to the different events during the fault, e.g., segment A indicates the pre-fault section and the fault-inception, segment B the fault, segment C opening of the circuit-breaker and segment D auto-reclosing of the circuit-breaker and system restore [1].

However, certain kinds of disturbance signals do not show distinct abrupt changes in the signal parameters, e.g., fault signals showing a gradual resistive decay. For those kinds of signals, application of the wavelet method using standard mother wavelets like Haar (Daubechies 1), Daubechies 4 [2], etc fail to achieve precise event-specific signal segmentations. One such example is shown in Figure 2, where the disturbance voltage recording comes from the DFR on a 765 kV long line with shunt reactors. The gradual drop in the voltage amplitude happens due to the release of the energy trapped by the R-L-C circuit, after the circuit-breakers are opened following the fault. As shown in Figure 2, the abrupt change detection-based segmentation in this type of cases causes confusing multiple close-spikes in the region of the fault-inception time (time when fault starts). In Figure 2, these close-spikes near the fault-inception are the segments A, B, C. In these cases, we propose to segment the signal into pre- and post-fault segments based on the fault-inception time. To accomplish that without causing multiple close spikes, a new adjustment technique to the Haar wavelet is proposed in the scope of this paper.

It is to be noted in this context that the current study is oriented towards augmenting the semi-automatic system to a fully automatic system. The former currently uses manual segmentation step with automatic post-processing. Therefore, in a semi-automatic case estimation of the fault-inception time is subject to human intervention. However, this is of critical importance for the futuristic fully automated system supposed to operate without any human intervention. In a fully automated system, the fault-inception time should be estimated as precisely as possible without any false positives as various important post-processing steps like synchronization, relay performance analysis etc depends on it. This is something that could not be performed precisely using the wavelet-based technique [1] for many signals like the one shown in Figure 2. And this is the motivation for the work presented in this paper.

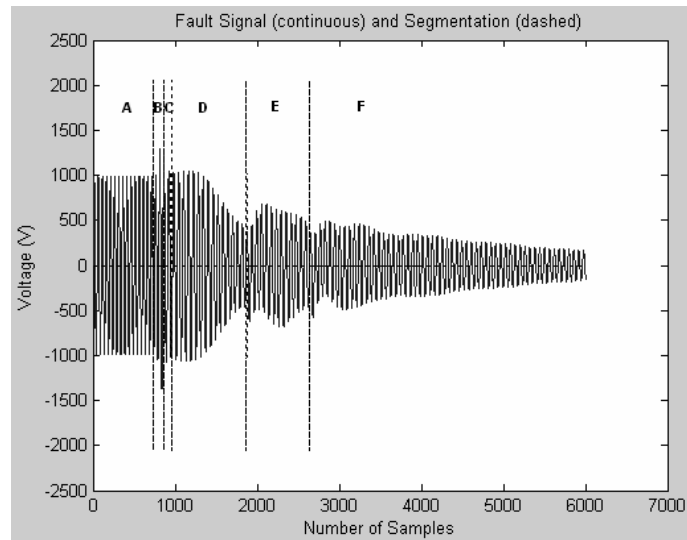

Figure 2.  Unsuccessful segmentation using the wavelet method for disturbance signal not showing distinct abrupt changes in the signal parameters

## 4     Adjusted Haar Wavelet

*4.1     Overview of Haar Wavelet*

'Haar' wavelet was first introduced by Alfred Haar in 1910. Interested reader can refer to the historical original German version of the paper in [6]. Haar wavelet is also referred to as Daubechies 1 [2] wavelet. The mathematical description of the Haar wavelet can be referred to in [2, 7, 8].

The scaling function $\phi(x)$ is defined as

$$\phi(x) = 1, \quad \text{if} \quad x \in [0,1],$$
$$\phi(x) = 0, \quad \text{if} \quad x \notin [0,1]. \tag{1}$$

The wavelet function $\psi(x)$ for this scaling function is defined as

$$\psi(x) = 1, \quad \text{if} \quad x \in [0, 0.5],$$
$$\psi(x) = -1, \quad \text{if} \quad x \in [0.5, 1], \qquad (2)$$
$$\psi(x) = 0, \quad \text{if} \quad x \notin [0, 1].$$

The scaling function and wavelet function for the Haar wavelet are shown in Figure 3 (a) & (b) respectively.

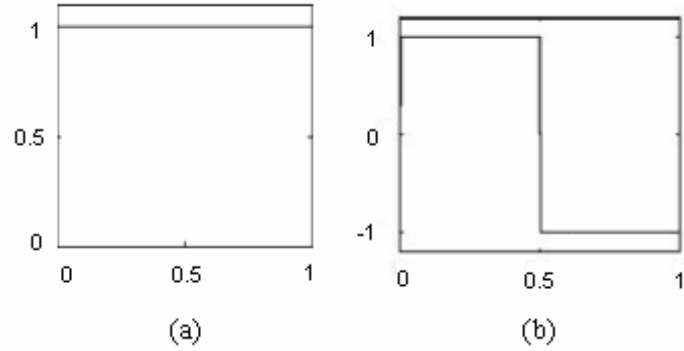

Figure 3. Haar wavelet: (a) Scaling Function, (b) Wavelet Function

*4.2   Adjustment to the Haar Wavelet*

In this section, we discuss the proposed adjusted Haar wavelet in terms of the key properties of the wavelets.

In general, the FIR (finite impulse response) scaling filter for the Haar wavelet looks like $h = 0.5[1 \ \ 1]$, where 0.5 is the normalization factor. However, the frequency domain response of the Haar wavelet scaling filter contains a lot of ripples which degrade its performance. As an adjustment and improvement of the characteristics of the Haar wavelet scaling filter in terms of reducing the ripples without violating the key wavelet properties, we propose to introduce 2*n* zeros (*n* is a positive integer) in the Haar wavelet scaling filter, keeping the first and last coefficients 1. Following the orthogonality property of the scaling filter, the filter length has to be even [7]. So, we have to introduce 2*n* adjusting zeros, *n* being the *adjustment* parameter. The original Haar wavelet corresponds to the condition $n = 0$. The introduced additional zeros in the filter kernel have zero coefficients. The scaling filter kernel for the adjustment parameter is shown below.

$$h = 0.5[1\ \ 1] \qquad \text{for } n = 0$$
$$h = 0.5[1\ 0\ 0\ 1] \qquad \text{for } n = 1 \qquad\qquad (3)$$
$$h = 0.5[1\ 0\ 0\ 0\ 0\ 1] \quad \text{for } n = 2$$
$$\vdots$$

It will be shown that the adjusting zeros improve the Haar wavelet characteristics, causing the strong ripples to die away quickly in the frequency domain. This is especially beneficial for our application. Also, it will be shown mathematically that the introduction of the adjusting zeros do not violate the key wavelet properties like the compact support, orthogonality and the perfect reconstruction. This will ensure the use of the adjusted Haar wavelet as a valid mother wavelet. We prove these results now.

*4.3    Compact Support*

LEMMA 4.1: The adjusted Haar wavelet with $2n$ adjusting zeros has compact support.

*Proof*:   The scaling filter of the Haar wavelet comes from an FIR filter, with finite length. The original Haar wavelet has the 'support' for the closed interval in continuous time for the range [0, 1], outside that the filter kernel *h* is zero. The words 'compact support' means that this closed set is bounded [7]. The wavelet is zero outside a bounded interval: compact support corresponds to FIR. The adjusted Haar wavelet scaling filter with the additional adjusting zeros also form an FIR filter kernel. And because the adjusting zeros have zero coefficients, we have the compact support for the closed interval [0,1] same as that of the original Haar wavelet. ■

*4.4    Orthogonality*

Real vectors are orthogonal (perpendicular) when $\mathbf{x}.\mathbf{y} = 0$. Real functions are orthogonal when $\int X(\omega)Y(\omega)d\omega = 0$. If the vectors or the functions are complex, we have to consider complex conjugates of one vector or one function. The discrete analogue of an orthonormal transform is a square matrix with orthonormal columns [7]. This is an 'orthogonal' matrix if real, a 'unitary' matrix if complex.

The orthogonal filter bank comes from an orthogonal matrix

$$A_t^T A_t = I \quad \text{and} \quad A_t A_t^T = I . \tag{4}$$

LEMMA 4.2: The adjusted Haar scaling filter with 2$n$ adjusting zeros is a symmetric, orthogonal FIR filter.

*Proof*: An FIR filter $H(z)$ is symmetric when $z^N H(z) = H(z^{-1})$. $N$ is odd for orthogonality and the filter length must be even [7]. The adjusted Haar scaling filter with two nonzero coefficients and 2$n$ adjusting zeros with zero coefficients form a filter kernel with $N = 2n+1$. If this is a symmetric filter, it has a length of $2n+2$ and has the form $(h(0), h(1), h(2),..., h(n), h(n),..., h(2), h(1), h(0))$ and by convention, $h(0)$ is the first nonzero coefficient. This vector must be orthogonal to all its double shifts [7]. The inner product with its shift by 1 must be $2h(0)h(1) = 0$; so $h(1) = 0$. Then the inner product with its shift by 2 gives $2h(0)h(2) = 0$; so $h(2) = 0$. Continuing like this, the inner product with its shift by $n$ gives $2h(0)h(n) = 0$; so $h(n) = 0$. So, the only nonzero coefficient for the symmetric, orthogonal filter is the $h(0)$ at both ends of the filter. Therefore the adjusted Haar scaling filter with two nonzero coefficients (equal to 1) at both ends and 2$n$ adjusting zeros with zero coefficients embedded in between form a symmetric, orthogonal FIR filter kernel. ■

*4.5 Perfect Reconstruction*

The Perfect Reconstruction condition for a lowpass filter $P_0(z)$ is

$$P_0(z) - P_0(-z) = 2z^{-l} . \tag{5}$$

Equation (5) can be simplified as discussed in [7], so that the perfect reconstruction condition states that the filter $P(z)$ must be a 'halfband filter' [7], so that

$$P(z) + P(-z) = 2 . \tag{6}$$

LEMMA 4.3: Introduction of the 2*n* adjusting zeros to the Haar wavelet scaling filter satisfies the perfect reconstruction condition.

*Proof*: The original Haar wavelet scaling filter has the form $h = [1\ 1]$, i.e., $H_0(\omega) = (1 + e^{-j\omega})$ and $H_0(z) = (1 + z^{-1})$. Introduction of the 2*n* adjusting zeros gives the scaling filter as $h = [1\ 0\ 0\ ...\ 0\ 0\ 1]$, i.e., $H_0(\omega) = (1 + e^{-j(2n+1)\omega})$ and $H_0(z) = (1 + z^{-(2n+1)})$, where *n* is a positive integer. The original Haar wavelet corresponds to $n = 0$. So, as per the perfect reconstruction condition shown in (6), for the adjusted Haar wavelet scaling filter we get,

$$H_0(z) + H_0(-z) = (1 + z^{-(2n+1)}) + (1 - z^{-(2n+1)}) = 2 . \tag{7}$$

This completes the proof. ∎

*4.6   Adjusted Scaling Function*

Following LEMMA 4.1, 4.2 and 4.3, we have established that the adjusted Haar wavelet scaling filter, with 2*n* adjusting zeros, satisfies the key wavelet properties like the compact support, orthogonality and the perfect reconstruction.

In the frequency domain, the adjusted Haar wavelet scaling filter with 2*n* adjusting zeros is given as

$$H_0(\omega) = \frac{1}{2}(1 + e^{-j(2n+1)\omega}). \tag{8}$$

Obviously, the function $H_0(\omega)$ in (8) is lowpass with $H_0(0) = 1$ and $H_0(\pi) = 0$. The formula for the Fourier transform of the scaling function based on the lowpass filter is given below as discussed by Qian [8].

$$\Phi(\omega) = H_0\left(\tfrac{\omega}{2}\right)\Phi\left(\tfrac{\omega}{2}\right) = H_0\left(\tfrac{\omega}{2}\right)H_0\left(\tfrac{\omega}{4}\right)\Phi\left(\tfrac{\omega}{4}\right) = \prod_{k=1}^{\infty} H_0\left(\frac{\omega}{2^k}\right)\Phi(0). \tag{9}$$

Using (8) & (9), the Fourier transform of the adjusted Haar wavelet scaling function is

$$\Phi(\omega) = \prod_{k=1}^{\infty} H_0\left(\frac{\omega}{2^k}\right) = \prod_{k=1}^{\infty} \frac{1}{2}\left(1 + e^{-j(2n+1)\frac{\omega}{2^k}}\right). \tag{10}$$

So,

$$\Phi(\omega) = \prod_{k=1}^{\infty} e^{-j(2n+1)\frac{\omega}{2^{k+1}}} \frac{1}{2}\left(e^{j(2n+1)\frac{\omega}{2^{k+1}}} + e^{-j(2n+1)\frac{\omega}{2^{k+1}}}\right) \tag{11}$$

$$= \prod_{k=1}^{\infty} e^{-j(2n+1)\frac{\omega}{2^{k+1}}} \cos\left(\frac{(2n+1)\omega}{2^{k+1}}\right)$$

$$= \exp\left\{-j(2n+1)\omega \sum_{k=1}^{\infty} \frac{1}{2^{k+1}}\right\} \prod_{k=1}^{\infty} \cos\left(\frac{(2n+1)\omega}{2^{k+1}}\right)$$

$$= e^{-j(2n+1)\frac{\omega}{2}} \prod_{k=1}^{\infty} \cos\left(\frac{(2n+1)\omega}{2^{k+1}}\right).$$

As per formula 1.439, p. 38 given by Gradshteyn and Ryshik [9],

$$\prod_{k=1}^{\infty} \cos\left(\frac{\omega}{2^{k+1}}\right) = \frac{\sin(\omega/2)}{\omega/2}, \tag{12}$$

we get the adjusted scaling function as,

$$\Phi(\omega) = e^{-j(2n+1)\frac{\omega}{2}} \frac{\sin((2n+1)\omega/2)}{(2n+1)\omega/2}. \tag{13}$$

Figure 4, 5, 6 show the pole-zero plots of the adjusted Haar wavelet scaling filters for $n = 0, 1, 2$ respectively. For the Haar wavelet scaling filter which is an FIR one, there are only the zeros (indicated

by the small circles in Figure 4–6), no poles. It is to be noted that the original Haar wavelet scaling filter corresponds to $n = 0$, and we introduce additional complex conjugate pairs of zeros for each $n > 0$.

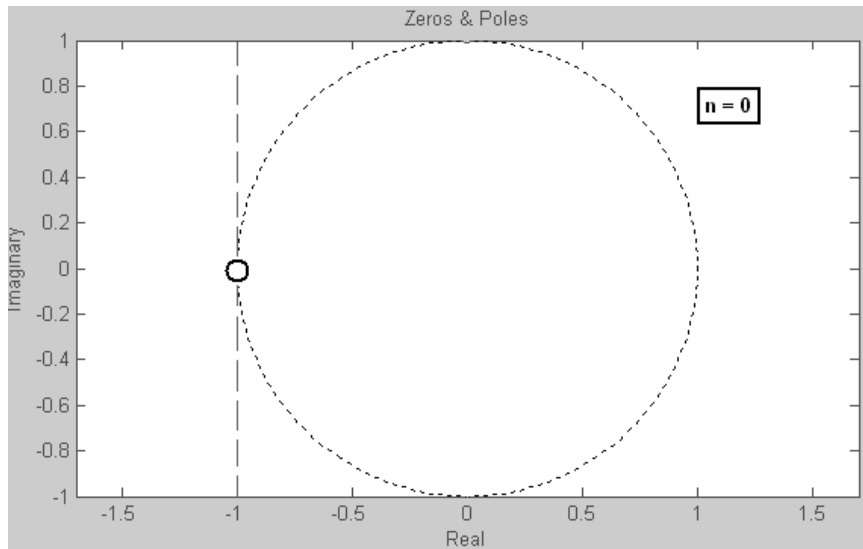

Figure 4. Pole-Zero plot of the adjusted Haar wavelet scaling filter, for *n*=0, which corresponds to the original Haar wavelet

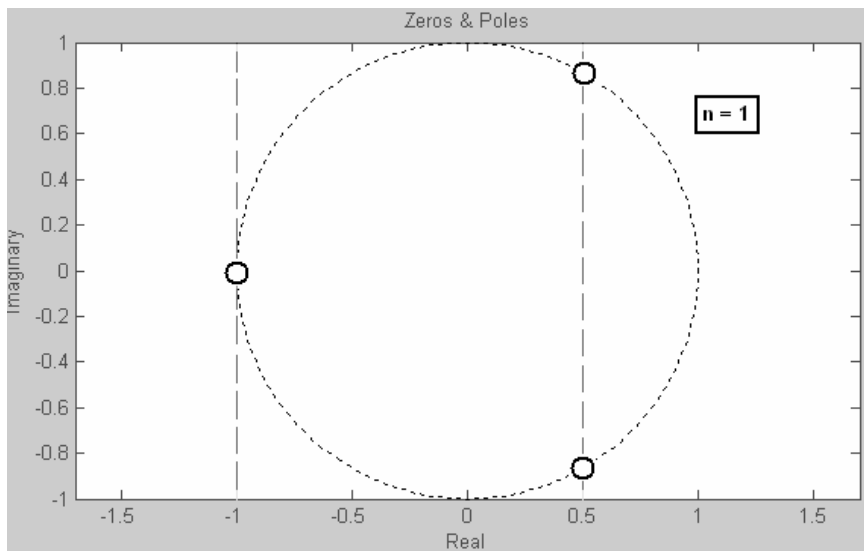

Figure 5. Pole-Zero plot of the adjusted Haar wavelet scaling filter with adjusting zeros, for *n*=1, i.e., one pair of complex conjugate zeros

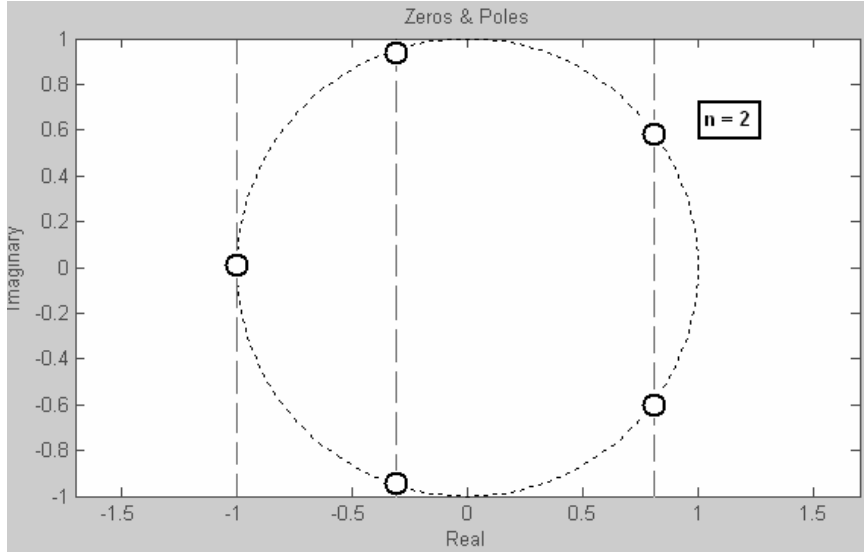

Figure 6. Pole-Zero plot of the adjusted Haar wavelet scaling filter with adjusting zeros, for *n*=2, i.e., two pairs of complex conjugate zeros

*4.7    Adjusted Wavelet Function*

To compute orthogonal mother wavelets from the lowpass filter $H_0(\omega)$, we need another function $H_1(\omega)$ such that

$$H_0(\omega)H_1^*(\omega) + H_0(\omega+\pi)H_1^*(\omega+\pi) = 0 . \tag{14}$$

This is condition for the quadrature mirror filter [7, 8]. One solution of (14) is

$$H_1(\omega) = -e^{-j\omega}H_0^*(\omega+\pi) . \tag{15}$$

Substituting $H_0(0) = 1$ and $H_0(\pi) = 0$ into (15) yields $H_1(0) = 0$ and $H_1(\pi) = 1$, respectively. This means that $H_1(\omega)$ in (15) is a highpass filter. So, for the adjusted Haar wavelet scaling (lowpass) filter $H_0(\omega)$ [see (8)], the highpass filter $H_1(\omega)$ is given by

$$H_1(\omega) = -e^{-j\omega}H_0^*(\omega+\pi) = -e^{-j\omega}\frac{1}{2}\left(1 - e^{j(2n+1)\omega}\right) \tag{16}$$

$$= \frac{1}{2}\left(e^{j2n\omega} - e^{-j\omega}\right).$$

Obviously, $H_0(\omega)$ and $H_1(\omega)$ constitute quadrature mirror filters, specified by (14). We can compute the Fourier transform of the wavelet function as discussed in [7], by

$$\Psi(\omega) = H_1\left(\frac{\omega}{2}\right)\Phi\left(\frac{\omega}{2}\right). \tag{17}$$

Now, we establish the main result, which is as follows.

THEOREM 4.4: Introduction of the $2n$ adjusting zeros to the Haar wavelet scaling filter improves the frequency characteristics of the adjusted wavelet function by an order of $2n+1$.

*Proof*: By (17), the Fourier transform of the adjusted wavelet function is [using (13) & (16)]

$$\Psi(\omega) = \frac{1}{2}\left(e^{jn\omega} - e^{-j\frac{\omega}{2}}\right) e^{-j(2n+1)\frac{\omega}{4}} \frac{\sin((2n+1)\omega/4)}{(2n+1)\omega/4} \tag{18}$$

$$= \frac{1}{2}\left(e^{j(2n-1)\frac{\omega}{4}} - e^{-j(2n+3)\frac{\omega}{4}}\right)\frac{\sin((2n+1)\omega/4)}{(2n+1)\omega/4}.$$

The magnitude is

$$|\Psi(\omega)| = \sqrt{2\{1-\cos((4n+2)\omega/4)\}\left\{\frac{2\sin((2n+1)\omega/4)}{(2n+1)\omega}\right\}^2}$$

$$= \sqrt{\{2\sin((2n+1)\omega/4)\}^2 \left\{\frac{2\sin((2n+1)\omega/4)}{(2n+1)\omega}\right\}^2}$$

$$= \frac{\{\sin((2n+1)\omega/4)\}^2}{|(2n+1)\omega/4|} < \frac{4}{|(2n+1)\omega|}. \tag{19}$$

The factor 2*n*+1 in the denominator of (19) improves the frequency characteristics of the adjusted Haar wavelet function, by decreasing the ripples (as *n*>0). This completes the proof. ∎

Figure 7 to 9 illustrate the proof above.

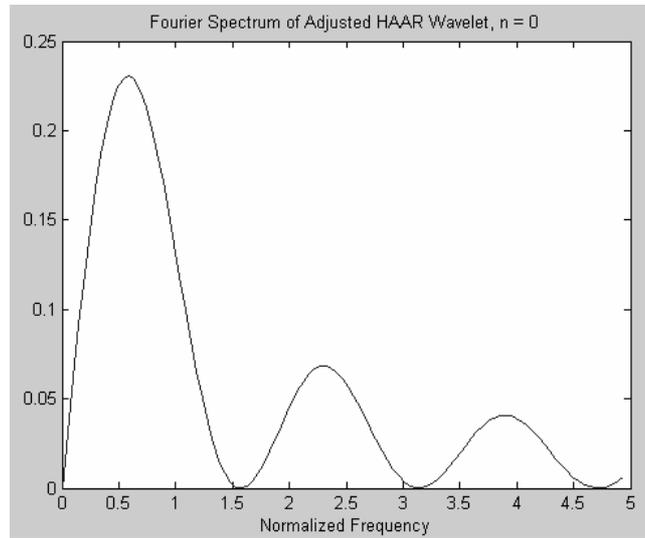

Figure 7. The Fourier spectrum of the adjusted Haar wavelet, for *n*=0, which corresponds to the original Haar wavelet

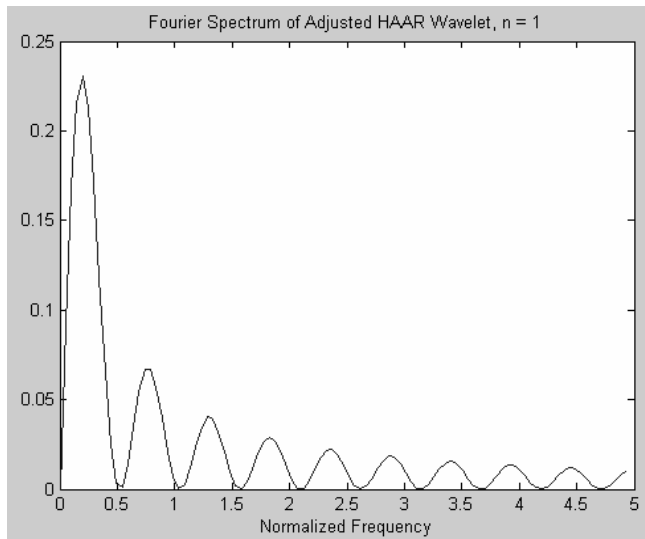

Figure 8. The Fourier spectrum of the adjusted Haar wavelet, for *n*=1, which decreases the strong ripples of the original Haar wavelet

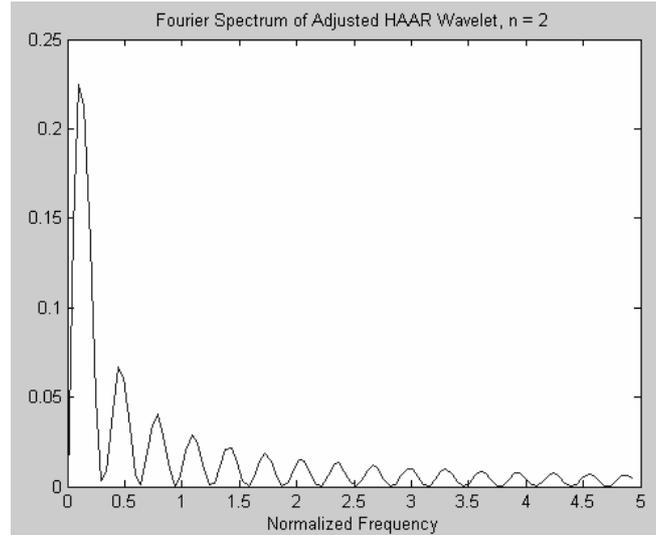

Figure 9.  The Fourier spectrum of the adjusted Haar wavelet, for *n*=2, which further decreases the strong ripples of the original Haar wavelet

The ripple is measured by the difference of the heights of the main (first) lobe and the secondary (second onwards) lobes in the magnitude plot. It is to be noted in Figure 7 to 9, that the aforesaid differences of heights increase gradually for a specific frequency band. This indicates improvement of the ripple as the ripples die away gradually more quickly from Figure 7 to 9.

Theoretically, larger the *n*, faster the ripples die away (see (19)). However, we cannot choose *n* arbitrarily large as that would make the filter response very slow. Therefore, from the practical point of view, we have to restrict the value of *n* to some finite values, typically between 1 to 4, depending on the application.

## 5     Application Results

After normalizing the original fault signal using its mean value, it is transformed into the smoothed and detailed version using the wavelet transform, with the adjusted Haar wavelet as the mother wavelet. Adjustment parameter *n*=2 is applied, i.e., 4 adjusting zeros are included in the adjusting Haar wavelet scaling filter. Then the threshold method [1] is applied on the detailed version to determine the change time-instants. Both the standard and the adjusted Haar wavelet use the same '*universal threshold*' of Donoho and Johnstone [10] to a first order of approximation. The universal threshold $T$ is given by

$$T = \sigma\sqrt{2\log_e n}, \qquad (20)$$

where $\sigma$ is the median absolute deviation of the wavelet coefficients [1] and $n$ is the number of samples of the wavelet coefficients.

This is followed by the smoothing filter operations [1] to indicate the change time-instants as unit impulses. MATLAB® with Wavelet toolbox [11] has been used for implementing the application. Figure 10 (a, b) to 13 (a, b) show the comparative results of the application of the original Haar wavelet (plot a) and the adjusted Haar wavelet (plot b) on the fault signals, sampled at a sampling frequency of 2.5 kHz [3], obtained from the DFRs of the power utility in South Africa, Eskom, during various disturbances. For these special disturbance signals, the original Haar wavelet fails to achieve correct segmentation whereas the adjusted Haar wavelet correctly segments the fault signals into pre- and post-fault segments shown as A and B respectively in the plot (b) of Figure 10 to 13. Figure 10 and 11 show the segmentation of the voltage waveforms during phase-to-ground faults, while Figure 12 and 13 show the segmentation of the voltage waveforms during phase-to-phase faults.

It is to be noted in Figure 10 to 13, in case of plot (a) that the segmentations using the standard Haar wavelet might seem performing better. However, the goal is to provide event-specific segmentations like the one shown in Figure 1. But the additional segmentations in the plot (a) of Figure 10 to 13 do not provide any additional event-specific information. Instead, these additional segmentations cause false alarms for the fault-inception instant which is very critical for further operation. Further operations like synchronization [12] depend on the fault-inception timing. So, for these kinds of signals which do not show distinct abrupt changes, adjusted Haar wavelet-based segmentation (plot b, Figure 10 to 13) correctly estimates the fault-inception instant removing the confusing false alarms.

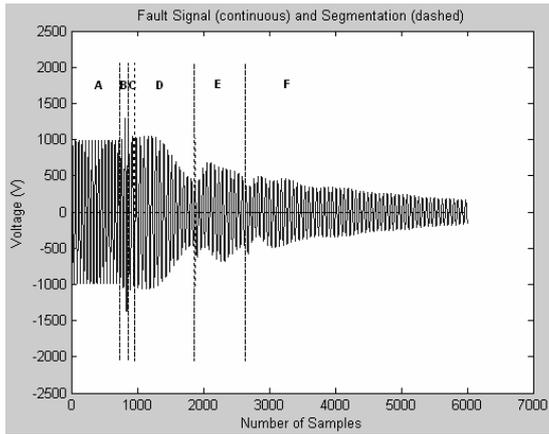 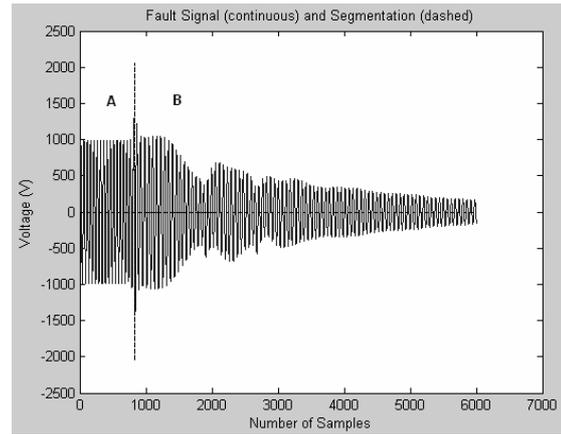

(a)                  (b)

Figure 10. Segmentation of the voltage waveform during a phase-to-ground fault, (a): using the standard Haar wavelet, (b): using the adjusted Haar wavelet

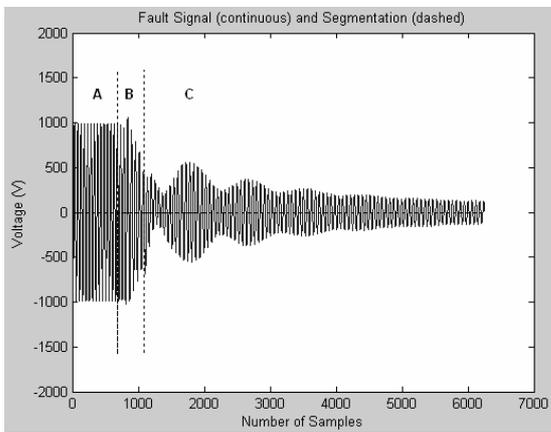 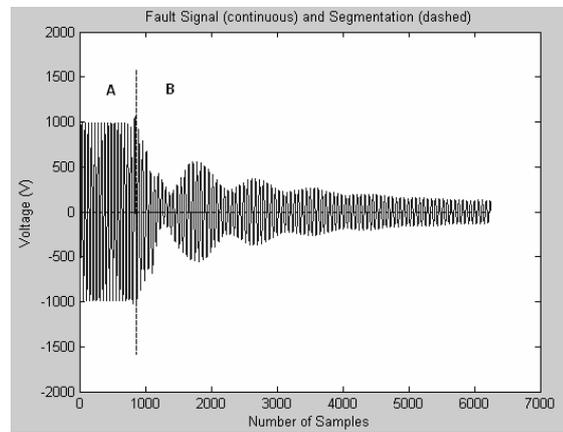

(a)                  (b)

Figure 11. Segmentation of the voltage waveform during a phase-to-ground fault, (a): using the standard Haar wavelet, (b): using the adjusted Haar wavelet

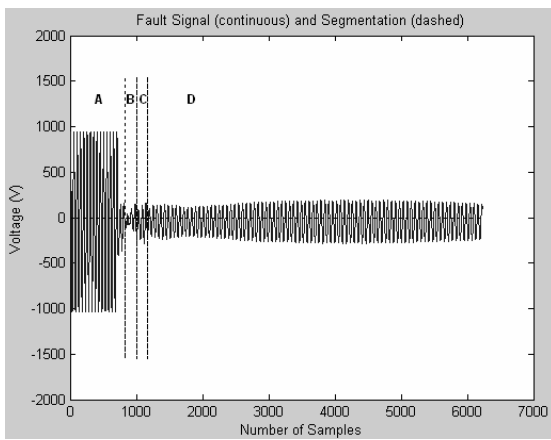 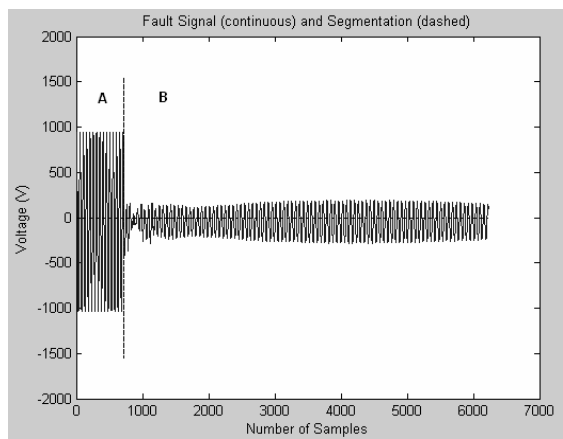

(a)                  (b)

Figure 12. Segmentation of the voltage waveform during a phase-to-phase fault, (a): using the standard Haar wavelet, (b): using the adjusted Haar wavelet

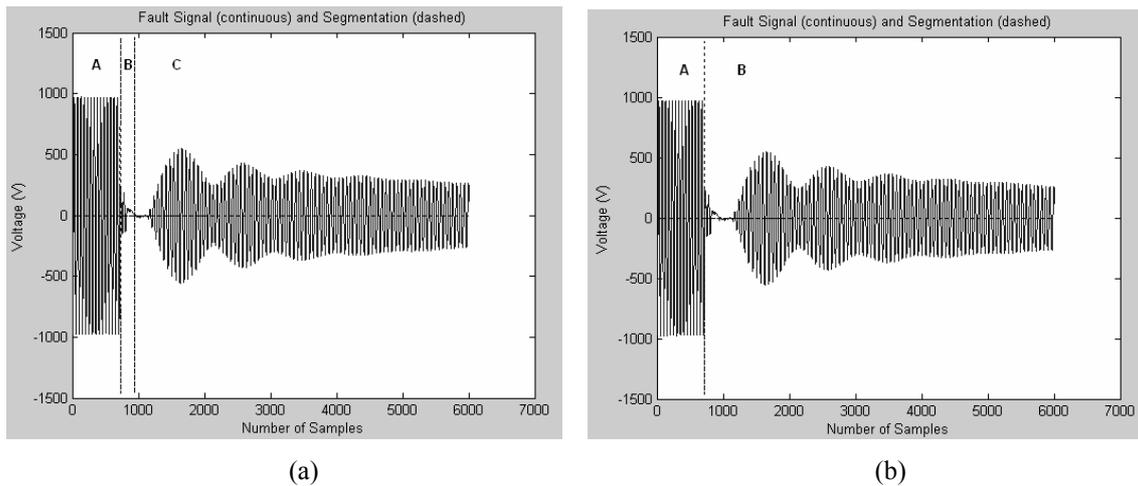

(a)                                                  (b)

Figure 13. Segmentation of the voltage waveform during a phase-to-phase fault, (a): using the standard Haar wavelet, (b): using the adjusted Haar wavelet

*5.1  Comments on Application Results*

The following comments can be cited on the application results.

- The proposed algorithm took an average computation time of 0.431 seconds. An Intel® Celeron® 1.9 GHz, 256 MB RAM computer was used for all the application tests using MATLAB® [11]. 250 critical disturbance records, unsuccessfully segmented using the Haar wavelet, have been tested. And 99% accuracy in correct segmentation, in terms of estimating the fault-inception time has been achieved.

- The authors proposed other methods for segmenting the signals without distinct abrupt changes, e.g., using the adaptive whitening filter along with the wavelet transform [1], recursive identification-based technique [13]. For estimating the fault-inception time for the signals without distinct changes, the accuracies of the adaptive whitening filter and the recursive identification method are 80% and 85% respectively, lower than the adjusted Haar wavelet-based method. Moreover, the whitening filter-based method [1] sometimes degrades the performance of the normal signals with distinct changes which does not occur in the case of adjusted Haar wavelet method. The comparatively slower recursive identification technique is mainly used for covering the cases of multiple close change spikes in signals with distinct changes not dealt correctly by the wavelet transform-based method [1]. So, in general, the adjusted Haar wavelet method is more robust for segmenting signals with and without distinct changes. Table 1 shows the comparison of the different methods.

Table 1

Comparison of the adjusted Haar wavelet method with other methods

| Method | No. of Signals Tested | Mean Time (sec) | % Accuracy for Fault-Inception Time | |
|---|---|---|---|---|
| | | | Without Changes | With Changes |
| Adjusted Haar Wavelet | 250 | 0.431 | 99 | 100 |
| Standard Wavelet | 250 | 0.431 | 70 | 99 |
| Adaptive Filter & Wavelet | 250 | 0.501 | 80 | 95 |
| Recursive Identification | 250 | 5.585 | 85 | 95 |

- Further processing, e.g., synchronization, analysis of relay performances, etc [12], of the fault signals without distinct abrupt changes depend on the post-fault segment and the fault-inception time-instant. The proposed adjusted Haar wavelet is very accurate in estimating the fault-inception time in these cases while other methods result in confusing false alarms.

# 6    Conclusion

Abrupt change detection using the wavelet transform and the threshold method is quite effective in segmenting the signals originated by power system disturbances into event-specific sections [1]. However, for certain kinds of disturbance signals not showing distinct abrupt changes in the signal parameters, standard mother wavelets e.g., Haar (Daubechies 1), Daubechies 4 fail to achieve correct event-specific signal segmentation. A novel technique for adjustment to the standard Haar wavelet is discussed in this paper. We propose to introduce $2n$ ($n$ is a positive integer) adjusting zeros in the Haar wavelet scaling filter, which improve the strong ripples in the frequency domain. It is also mathematically established that the adjusted Haar wavelet scaling filter, with 2*n* adjusting zeros, and the resulting adjusted wavelet function satisfy the key wavelet properties like the compact support, orthogonality and the perfect reconstruction. 250 critical disturbance records without distinct abrupt changes in the signal parameters have been tested with the adjusted Haar wavelet. And the accuracy rate in estimating the fault-inception time-instant improved significantly using the adjusted Haar wavelet compared to the standard mother wavelets or other adaptive filtering-based techniques.

**Acknowledgments**

This work was supported in part by the National Research Foundation (NRF), South Africa.

All real fault signal recordings were kindly provided by Eskom, South Africa.

**Vitae**

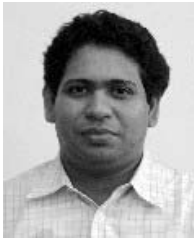

**Abhisek Ukil** received the B.E. degree from the Jadavpur University, Calcutta, India, in 2000 and the M.Sc. degree from the University of Applied Sciences, South Westphalia, Soest, Germany, and the University of Bolton, Bolton, UK, in 2004. He received the Dr. Tech. degree from the Tshwane University of Technology, Pretoria, South Africa in 2006. Currently, he is a research scientist at ABB Corporate Research Center in Baden, Switzerland. His research interests include applied signal processing, machine learning, intelligent systems, power systems.

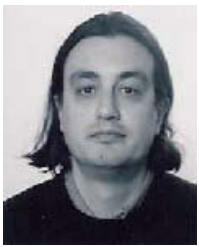

**Rastko Živanović** received the Dipl.Ing. and M.Sc. degrees from the University of Belgrade, Belgrade, Serbia, in 1987 and 1991, and the Ph.D. degree from the University of Cape Town, Cape Town, South Africa, in 1997. Previously, he was with Tshwane University of Technology, Pretoria, South Africa. Currently, he is senior lecturer with the School of Electrical & Electronic Engineering at the University of Adelaide, Adelaide, Australia. His research interests include signal processing, power system protection.

**Figure Captions**

Figure 1. Abrupt change detection-based segmentation of the BLUE-phase voltage recording during a phase-to-ground fault

Figure 2. Unsuccessful segmentation using the wavelet method for disturbance signal not showing distinct abrupt changes in the signal parameters

Figure 3. Haar wavelet: (a) Scaling Function, (b) Wavelet Function

Figure 4. Pole-Zero plot of the adjusted Haar wavelet scaling filter, for $n=0$, which corresponds to the original Haar wavelet

Figure 5. Pole-Zero plot of the adjusted Haar wavelet scaling filter with adjusting zeros, for $n=1$, i.e., one pair of complex conjugate zeros

Figure 6. Pole-Zero plot of the adjusted Haar wavelet scaling filter with adjusting zeros, for $n=2$, i.e., two pairs of complex conjugate zeros

Figure 7. The Fourier spectrum of the adjusted Haar wavelet, for $n=0$, which corresponds to the original Haar wavelet

Figure 8. The Fourier spectrum of the adjusted Haar wavelet, for $n=1$, which decreases the strong ripples of the original Haar wavelet

Figure 9. The Fourier spectrum of the adjusted Haar wavelet, for $n=2$, which further decreases the strong ripples of the original Haar wavelet

Figure 10. Segmentation of the voltage waveform during a phase-to-ground fault, (a): using the standard Haar wavelet, (b): using the adjusted Haar wavelet

Figure 11. Segmentation of the voltage waveform during a phase-to-ground fault, (a): using the standard Haar wavelet, (b): using the adjusted Haar wavelet

Figure 12. Segmentation of the voltage waveform during a phase-to-phase fault, (a): using the standard Haar wavelet, (b): using the adjusted Haar wavelet

Figure 13. Segmentation of the voltage waveform during a phase-to-phase fault, (a): using the standard Haar wavelet, (b): using the adjusted Haar wavelet

**Table Captions**

Table 1. Comparison of adjusted Haar wavelet method with other methods